\documentstyle[prl,aps,epsf,preprint]{revtex}
\begin{document}
\title{NMR relaxation rates in spin-1/2 antiferromagnetic chains.}
\author{Victor Barzykin}
\address{National High Magnetic Field Laboratory, 
Florida State University,\\
1800 E. Paul Dirac Dr., Tallahassee, Florida 32310}
\maketitle
\begin{abstract}
We consider the low-temperature behavior of  
NMR relaxation rates $T_1$ and $T_{2G}$ in the spin-$1/2$ 
antiferromagnetic chain. We find that $T_1 \propto const$ 
and $T_{2G} \propto \sqrt{T}$, with logarithmic corrections.
We determine both constants and logarithmic terms by matching 
perturbative RG with exact results, so that the final expressions 
for the relaxation rates do not contain any free parameters.
Our theoretical results are in excellent agreement with
NMR experiments in Sr$_2$CuO$_3$.
\end{abstract}
\newpage
There has been significant interest in the low-dimensional quantum spin systems
and quantum critical phenomena in recent years due to the discovery of 
high-temperature superconductivity in layered cuprates. 
Nuclear magnetic resonance (NMR)
has been a powerful tool in studying the spin dynamics of the 
cuprate compounds\cite{BP}. The measurements
of the longitudinal relaxation rate $1/T_1$, the  spin echo
decay rate $1/T_{2G}$, and the Knight shift over a wide temperature
range define the spectrum of antiferromagnetic fluctuations
in the high-T$_c$ cuprates\cite{ZBP}. These NMR experiments have been 
the basis of the phenomenological spin-fluctuation model\cite{MMP}, which
exhibits quantum critical behavior\cite{BP}, because the system always
remains close to the antiferromagnetic instability.

Quantum spin-$1/2$ spin chain is another example
of a quantum critical system with novel properties. The physical
properties of this system are well understood theoretically.
The XXZ spin-$1/2$ quantum spin chain is described by the Hamiltonian:
\begin{equation}
H = J \sum_i [S_i^xS_{i+1}^x + S_i^yS_{i+1}^y + \gamma S_i^zS_{i+1}^z].
\end{equation}
This model exhibits quantum critical behavior for $-1 \leq \gamma \leq 1$, with
asymptotic correlation functions vanishing with distance as a power law.
Inverse temperature $1/T$ acts essentially as a finite system size,
so that the power law for correllation function crosses over to
exponential decay at distances $r > \xi_T \simeq \hbar c/T$ 
For $|\gamma| > 1$ the spin-$1/2$ chain is equivalent to a model of massive 
free fermions, and the correlation functions decay exponentially with distance.
Exactly at the Heisenberg point, $\gamma =1$, logarithmic corrections
to the correlation functions appear as a result of the presence of 
the leading marginally irrelevant operator. 

Logarithmic corrections in the SU(2)-invariant models are well known. 
For example, a calculation of the logarithmic corrections up to two 
loops was done for the fermion model with backward scattering\cite{Menyhard},
the sine Gordon model\cite{Amit}, and the SU(2) Gross-Neveu model\cite{Wetzel}.
Logarithms appear in every physical property of the spin-$1/2$ Heisenberg chain.
For example, $1/\ln(H)$ dependence of magnetization on magnetic field 
was found in Ref. \onlinecite{Babujian} and in 
Refs. \onlinecite{Schlottmann1,Schlottmann2}. The 
temperature-dependent $1/\ln(T)$ corrections to the bulk spin 
susceptibility\cite{EAT}. Finite-size scaling corrections to the staggered
spin susceptibility were found in Ref. \onlinecite{BA}.
In what follows we determine the logarithmic behavior of the nuclear magnetic
resonance (NMR) relaxation rates $T_1$ and $T_{2G}$.

Nuclear spins probe local spin environment. The Knight shift provides a
measure of the uniform magnetic susceptibility at a particular nuclear
site, while the experiments on the spin-lattice relaxation rate yield
information on the imaginary and real parts of the dynamic 
spin susceptibility $\chi({\bf q}, \omega)$.
The analysis of the NMR experiments begins with the
magnetic hyperfine hamiltonian, which couples nuclear spins and conduction
electron spins:
\begin{equation}
H_{HF} = \sum_{\alpha,i,j} A^{ij}_{\alpha}I_{i \alpha} S_{j \alpha},
\end{equation}
$I$ is the nuclear spin, $S$ is the electron spin, $\alpha$ enumerates
spin projections for sites $i$ and $j$ . The following 
expressions can be obtained\cite{slichter} for $T_1$ and $T_{2G}$:
\begin{eqnarray}
{1 \over T_1} &=& {2 k_B T \over  \hbar^2} \int {dq \over 2 \pi}
A_{\perp}^2(q){Im\chi(q,\omega_0) \over \omega_0} \\
\left({1 \over T_{2G}}\right)^2 &=& {0.68 \over 8 \hbar^2} \left[
\int {dq \over 2 \pi} 
A_{\parallel}^4(q) \chi^2(q) - \left\{\int {dq \over 2 \pi}
A_{\parallel}^2(q) \chi(q) \right\}^2 \right].
\end{eqnarray}
Here $A_{\parallel}(q)$ and $A_{\perp}(q)$ are the hyperfine 
couplings parallel and perpendicular to the easy axis of the
crystal, $\omega_0$ is the nuclear resonance frequency, which is much
smaller than any other electron  energy scale. The magnetic 
field is directed along the $c$-axis.
The q-dependence is smooth and arises from appropriate form 
factors. The susceptibility $\chi$ should, in principle, include contributions
from both the uniform and staggered spin fluctuations. However, simple
power counting\cite{Sachdev} shows that the staggered component is dominant
at small $T$. Indeed, the contribution of the uniform component  scales
as  $1/T_1 \propto T$, $1/T_{2G} \propto T^0$, while for the staggered
component $1/T_1 \propto T^0$, $1/T_{2G} \propto T^{-1/2}$.

For the purpose of comparison of theory with experiment, it is 
convenient to define normalized dimensionless NMR relaxation rates:
\begin{eqnarray}
\left({1 \over T_1}\right)_{norm.} &=& {\hbar J \over A_{\perp}^2 T_1} \\
\left({\sqrt{2} \over T_{2G}}\right)_{norm.} &=& \left(k_B T \over pJ\right)^{1/2}
{\hbar J \over A_{\parallel}^2(\pi) T_{2G}}
\end{eqnarray}

The calculation of the NMR relaxation rates are based on the continuum limit
bozonised approximation to the Heisenberg model. The Hamiltonian density
can be written as:
\begin{equation}
\label{hamilt}
H = H_0 - 2 \pi g \vec{J}_L \vec{J}_R.
\end{equation}
Here $H_0$ is the Hamiltonian density for a free boson of compactification
radius $R=1/\sqrt{2 \pi}$, $\vec{J}_L$ and $\vec{J}_R$ are the left and
right moving currents. The coupling constant obeys the renormalization group
equations known for the Kosterlitz-Thouless or the Kondo problem with 
ferromagnetic interactions\cite{BAK}:
\begin{equation}
\beta \equiv dg/d\ln\Lambda = - g^2 - {1 \over 2} g^3,
\end{equation}
as the energy cutoff $\Lambda$ varies.

The low-temperature asymptotics for NMR relaxation rates are determined
by the staggered component of the dynamic spin correlator, which
has the following general form\cite{barzykin}:
\begin{equation}
\label{sol}
\chi_s(r, t, T, g_0) = D Z(g(T)) F[g(T), r T/c, t T].
\label{stag}
\end{equation}
Here $Z(g(T))$ is a cutoff-dependent renormalization factor, $C$
is a nonuniversal constant. This scaling form can be easily obtained from
the Callan-Symanzik Renormalization Group equations:
\begin{equation}
\label{RG}
[- \partial/\partial \ln{T}
+\beta(g)\partial/\partial g +
2 \gamma(g)] \chi_s(r,r T/c,t T,  g) = 0,
\end{equation}
where $\beta(g)$ is the beta function for the
coupling constant $g$ in Eq.(\ref{hamilt}):
\begin{equation}
{d g \over d \ln{T}} = - \beta(g) = g^2 - g^3/2 + O(g^4)
\label{beta}
\end{equation}
and $\gamma(g) \simeq 1/2 - g/4 + O(g^2)$ is the anomalous dimension.  
In Eq. (\ref{RG}) the $T$-derivative acts only on the first argument 
of $\chi_s$; $rT$ and $t T $ are
held fixed. The solution of Eq.(\ref{RG}) can then be written in the
form Eq. (\ref{sol}) with
\begin{eqnarray}
D & \propto & exp\left(2 \int_{0}^{g_0} {\gamma[g'] \over \beta(g')} d g' \right) \\
Z(g) &=& exp\left(- 2 \int_{0}^{g(T)} {\gamma[g'] \over \beta(g')} d g' \right),
\end{eqnarray}
where $g_0 \equiv g(\Lambda)$ is the "bare"
coupling - the coupling at the
energy cutoff scale $\Lambda$.

The leading behavior of $Z(g)$ is easily seen from 
perturbative expressions for $\beta(g)$ and $\gamma(g)$ given
above\cite{Affleck}, $Z(g) \propto 1/\sqrt{g}$. 
Integrating Eq.(\ref{stag}) gives 
\begin{eqnarray}
{1 \over T_1} & \propto & Z(g(T)) (1 + a_1 g(T) + a_2 g(T)^2 + 
\cdots) \\ \nonumber
{1 \over T_{2G}} &\propto& {Z(g(T)) \over \sqrt{T}} 
(1 + b_1 g(T) + b_2 g(T)^2 + \cdots).
\end{eqnarray}
The non-universal constants $a_i$, $b_i$, the cutoff $\Lambda$ and 
common factor $D$ in these expressions can be fixed using
exact Bethe Ansatz results on the correlation 
functions. The value of $D=1/(2 \pi)^{3/2}$ was determined in 
Ref.\onlinecite{Lukyanov,Affleck1}.

A complete calculation of the NMR relaxation rates gives:
\begin{eqnarray} 
(1/T_1)_{norm} & = & 2 D \sqrt{\ln{\Lambda \over T} + {1 \over 2}
\ln\left(\ln{\Lambda \over T}\right)}\left(1 + O\left[{1 \over \ln^2{\Lambda \over T}}\right]\right) \\
(\sqrt{T}/T_{2G})_{norm} & = &  {\sqrt{I_0} D \over 4 \sqrt{\pi}}
\sqrt{\ln{\Lambda_1 \over T} + {1 \over 2} 
\ln\left(\ln{\Lambda_1 \over T}\right)} \left(1 +
O\left[{1 \over \ln^2{\Lambda_1 \over T}}\right]\right).
\end{eqnarray}
Here $D = 1/(2 \pi)^{3/2}$ is the non-universal amplitude, 
$C \simeq 0.5772157$ is the Euler's constant, while the integrals
$I_0$ and $I_1$ are given by
\begin{eqnarray}
I_0 & = & \int_0^{\infty} dx \left| {\Gamma\left({1 + ix \over 4}\right) \over
\Gamma\left({3+ix \over 4}\right)}\right|^4 \simeq 71.2766 \nonumber \\
I_1 &=&  \int_0^{\infty} dx \left| {\Gamma\left({1 + ix \over 4}\right) \over
\Gamma\left({3+ix \over 4}\right)}\right|^4 \times \nonumber \\
 & \times &Re\left[\Psi\left({1 + ix \over 4}\right) + \Psi\left({3 + ix \over 4}\right)\right] \simeq - 259.94,
\label{rates}
\end{eqnarray}
where $\Psi(x)$ is digamma function. The $1/\ln(\Lambda/T)$ term is
incorporated in the cutoff as in Ref\cite{barzykin}.
Thus, up to terms $O(1/\ln^2(\Lambda/T))$ the temperature dependence 
for $1/T_1$ or $\sqrt{T}/T_{2G}$ is actually given by the square root
of the $log$ and $loglog$ terms in the numerator of Eq.(\ref{rates}). 
The cutoff parameters $\Lambda$ can be determined using exact methods:
\begin{eqnarray}
\Lambda &=& 2 \sqrt{2 \pi} e^{C+1} J \simeq 24.27 J\\
\Lambda_1 &=&{\sqrt{2 \pi} e \over 8} e^{- I_1/2I_0} J \simeq 5.27 J
\end{eqnarray}

The
ratio of the relaxation rates, however, is only weakly temperature dependent,
\begin{equation}
\left(T_{2G} \over T_1 \sqrt{T}\right)_{norm} \simeq 1.680 
\left(1 + {0.7632 \over \ln(\Lambda/T)} \right) 
\end{equation}

 In summary, I have presented theoretical low temperature results
for NMR rates $T_1$ and $T_{2G}$.The temperature dependence of $1/T_1$ 
is logarithmic, which is expected in the presence of the leading irrelevant
operator. The main improvement over past theories is that the amplitude and 
the cutoffs were also calculated, so our expressions do not contain any 
free parameters. This has been a matter of confusion in the past\cite{takig}.
In particular, cutoffs have been fitted and used together with $1/log$ terms which 
redefine them. Presumably, this explains the difference of our cutoff 
$\Lambda$ with fitted value of $4.5J$ used in Ref. \onlinecite{takig}. Our 
theoretical results agree reasonably well with experimental data of 
Takigawa {\em et al.}\cite{takig} on Sr$_2$CuO$_3$, as shown in 
Fig.1 and Fig.2.

This work was supported by the National High Magnetic Field Laboratory through 
NSF cooperative agreement No. DMR-9527035 and the State of Florida.

\begin{figure}
\epsfxsize=7 in
\epsfbox{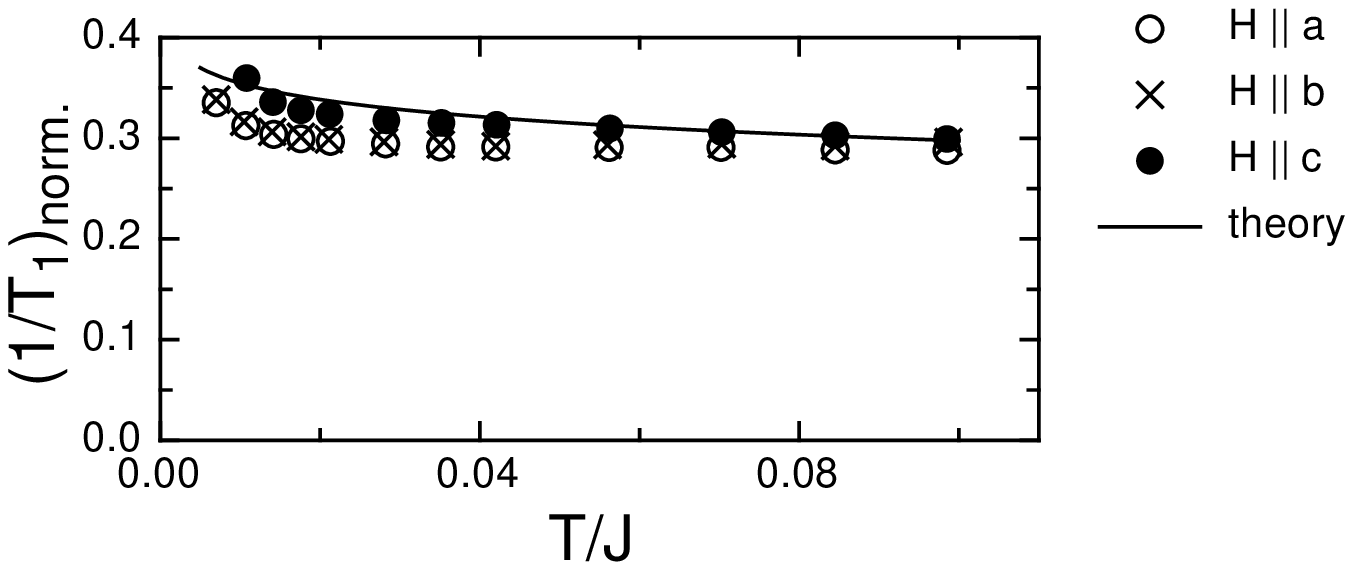}
\caption{NMR $1/T_1$ for magnetic fields along different crystal axes in Sr$_2$CuO$_3$ from 
Takigawa {\em et al.} 
\protect{\cite{takig}}
compared to our theoretical expression with no adjustable parameters. }
\end{figure}
\newpage
\begin{figure}
\epsfxsize=7 in
\epsfbox{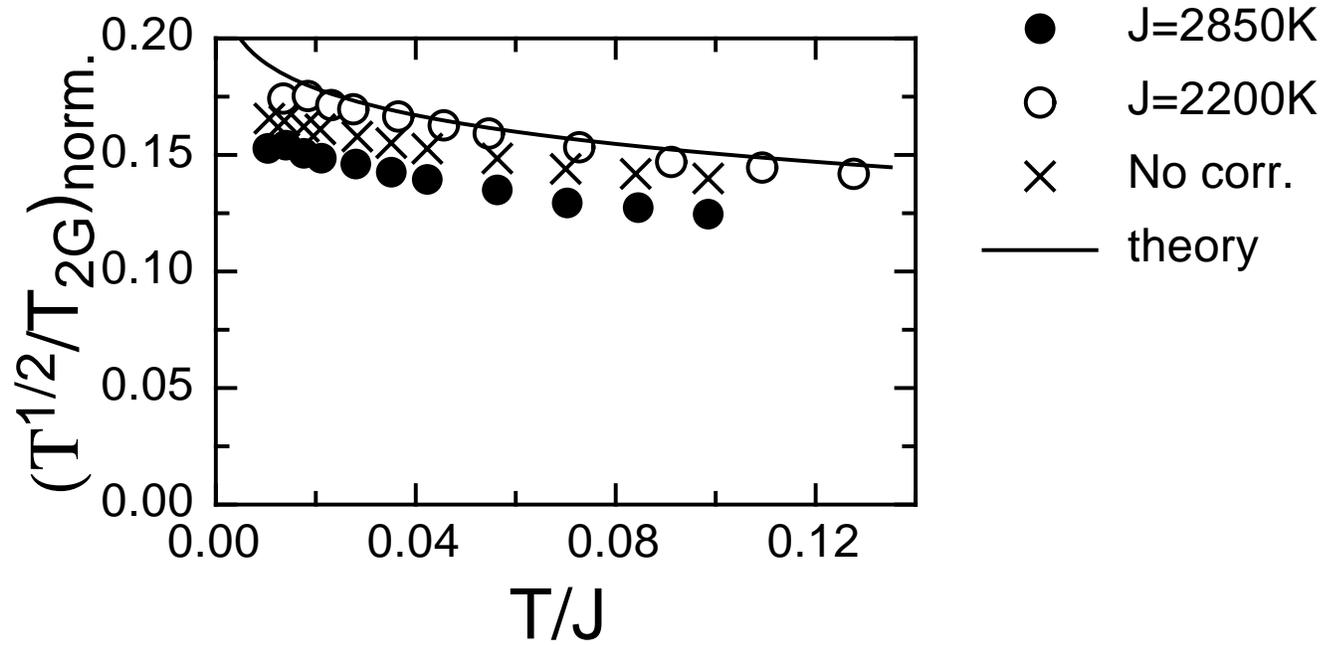}
\caption{NMR $T^{1/2}/T_{2G}$ from Takigawa {\em et al.} 
\protect{\cite{takig}} 
vs our low-temperature result. The data shows different assumptions for the value of
$J$ (fitted from the spin susceptibility data). The crosses are the data for $J=2850K$
not corrected for the $I_z$ fluctuations.}
\end{figure}
\end{document}